\newcommand{\be}{\begin{equation}}
\newcommand{\ee}{\end{equation}}
\newcommand{\ber}{\begin{eqnarray}}
\newcommand{\eer}{\end{eqnarray}}

\newcommand{\CL}{{\cal L}}
\newcommand{\Tr}{{\rm Tr}}

\newcommand{\diag}{{\rm diag}}
\newcommand\eqn[1]{\label{eq:#1}} 
 
\newcommand\eq[1]{Eq.~(\ref{eq:#1})}

\documentclass[twocolumn,showpacs,preprintnumbers,amsmath,amssymb]{revtex4}

\usepackage{psfrag}
\usepackage{epsfig}

\begin{document}
\tighten
\preprint{\vbox{\hbox{INT-PUB 04-10}}}
\bigskip
\title{New phases in CFL quark matter}
\author { Andrei Kryjevski\footnote{abk4@phys.washington.edu},  David B. Kaplan\footnote{dbkaplan@phys.washington.edu}
 } 
\affiliation{Institute for Nuclear Theory,  University of Washington, 
Seattle, WA 98195}
\author{Thomas Sch\"afer\footnote{thomas\_schaefer@ncsu.edu}}
\affiliation{Department of Physics, North Carolina State University, 
Raleigh, NC 27695 \\
and Riken-BNL Research Center, Brookhaven National Laboratory, Upton, NY 11793}

\date{\today}

\begin{abstract}
We consider $O(\alpha_s)$ corrections to the  squared masses of 
the pseudo-Goldstone excitations about the ground state of dense quark 
matter. We show that these contributions tend to destabilize the
vacuum, leading to a surprisingly complex phase structure   
for quark matter as a function of quark mass, even for small $\alpha_s$.  
In particular we find two new phases of CFL quark matter possibly relevant for the
real world, for which  $\bar\theta_{\rm QCD} = \pi/2$. 

\end{abstract}


\maketitle

The dependence of the QCD vacuum on the masses of the light quarks is
most efficiently analyzed by means of a chiral Lagrangian, which
allows one to systematically compute corrections to the chirally symmetric 
vacuum in a power expansion in the quark masses.  The utility of the chiral
Lagrangian approach is that it properly accounts for the lightest QCD
excitations, the pseudo Goldstone bosons (PGBs), whose masses vanish
in the chiral limit and which therefore have the most important
role in determining the vacuum structure for small quark mass.
In addition to allowing one to compute perturbative  
corrections to the vacuum structure, the chiral Lagrangian also allows one 
to  investigate  phase transitions at critical values of the quark
masses.  An example of such a phase transition was first discovered by Dashen
\cite{Dashen:1971et}, who found a line of phase transitions in the
$(m_u,m_d)$ plane corresponding to spontaneously broken CP symmetry,
where $m_{u,d}$ are the up and down quark masses respectively.  

In hadronic matter at nonzero baryon  density, more complicated 
phase transitions have been investigated using chiral Lagrangians, 
including both pion \cite{Migdal:1978az} and kaon
condensation \cite{Kaplan:1986yq,Nelson:1987dg}. More  recently, chiral
Lagrangians have been used to analyze the vacuum structure of dense 
quark matter near the chiral limit. It has been convincingly argued 
that for degenerate quarks such a system should be color
superconducting \cite{Bailin:1984bm}, 
and that for three massless flavors the color-flavor-locked (CFL) 
ground state is preferred \cite{Alford:1998mk,Schafer:1999fe,Evans:1999at}.  
Chiral perturbation theory can be used to describe the ground state and 
excitations of CFL matter away from the chiral limit
\cite{Casalbuoni:1999wu,Son:1999cm,Son:2000tu}. It is already known that 
the phase structure is  much more complicated than in the vacuum case.  
In particular, an analysis to leading order in both quark masses and 
the QCD coupling constant $\alpha_s$ reveals a phase transition to a 
kaon condensed phase along a line in the $m_s$-$m$ plane 
\cite{Schafer:2000ew,Bedaque:2001je,Kaplan:2001qk}, where $m_{s}$ is 
the strange quark mass, and for simplicity we are considering the
isospin limit, $m_u=m_d=m$. In this Letter we show that the leading 
order calculation does not capture the full complexity of the quark 
matter ground state, and that there exist new phases revealed at 
sub-leading order in $\alpha_s$  at small quark mass in the $m_s$-$m$ 
plane which could be relevant for the real world.

 The reason that $O(\alpha_s)$ effects are important is that the 
leading order result exhibits an accidental degeneracy which is 
only lifted at next-to-leading order. Once $O(\alpha_s)$ effects 
are included the meson masses receive negative corrections to
their squared masses which can be larger than the leading order
contributions, even for small $\alpha_s$.  In particular, a meson of
mass $M$ with
$M^2 \propto m^2$ at leading order can become destabilized by a
perturbative correction of the form $\delta M^2 \propto  -\alpha_s
m_s^2$, given that $m_s^2 \gg m^2$. Characteristic of these new
phases, which can coexist with kaon condensation, is that they break
an exact $Z_2$ symmetry (parity)  and that $\det{U}\ne 1$, where $U$ 
is the unitary matrix characterizing the quark condensate.  Such
phases appear to  
have a non-vanishing strong CP violating angle, $\bar \theta_{QCD}=\pi/2$.

In the CFL phase, attractive interactions near the (shared) Fermi surface
cause quarks of all three flavors to undergo BCS-like pairing. The 
resulting condensate takes the form
\be
\langle \psi_i^a C\gamma_5 \psi_j^b \rangle 
\propto \Delta_3\,  \epsilon^{a b x}\epsilon_{i j x} + 
  \Delta_6\,\left(\delta^a_i\delta^b_j + \delta^a_j\delta^b_i\right),
\eqn{cond}
\ee
where $\Delta_{3,6}$ are the pairing gaps in the antisymmetric 
color triplet and symmetric color sextet channels respectively. 
A perturbative calculation yields \cite{Schafer:1999fe}
\be
\Delta_6^2 = \alpha_s\frac{( \ln 2)^2\,}{162\pi}\Delta_3^2\ .
\eqn{dsix}
\ee
The gap $\Delta_3$ is larger than $\Delta_6$ because  one-gluon
exchange is attractive 
 in the triplet channel but repulsive in the sextet 
channel; the radiative generation of nonzero $\Delta_6$ does not
affect the symmetries of the CFL phase.  The condensate  spontaneously 
breaks the $U(1)_B$ baryon number symmetry, as well as the  
$SU(3)_{c} \times SU(3)_L \times SU(3)_R$ symmetry  down to 
the diagonal subgroup, $SU(3)_{c+L+R}$ where $SU(3)_c$ refers 
to the color gauge group.  As a result the gluons become massive, 
and there appears a nonet of Goldstone bosons analogous to the 
pseudoscalar meson nonet in the vacuum (in addition to the 
superfluid mode from the breaking of $U(1)_B$). Unlike in the 
vacuum, instantons are suppressed at high density, and the entire 
nonet is expected to be light.

 In the CFL phase low energy excitations may therefore be parametrized by
$\Sigma =e^{2i\pi/f_\pi}$, $V=e^{4i \eta'/ f_{\eta'}}$, and
$B=e^{i \beta/f_{B}}$ with $\pi = \pi^a T^a$, the $T^a$ being
$SU(3)$ generators.  The $\pi_a$ and $\eta'$  form the 
pseudo-scalar nonet, while  $\beta$ (which we shall ignore for the
remainder of this Letter) is the Goldstone boson of broken baryon number. 
Under the original symmetry of $SU(3)_{L}\times SU(3)_{R}\times U(1)_{B}
\times U(1)_{A}$ the $\Sigma$ field transforms as $(3,\bar 3)_{(0,0)}$,
$V$ transforms as  $(1,1)_{(0,4)}$, where we have assigned axial charges 
of $+1$ and $-1$ to the right- and left-handed quarks respectively. A 
gauge invariant order parameter for the system is (schematically) 
$\langle \bar q_L \bar q_L q_R q_R\rangle$ whose orientation in 
$SU(3)_L\times SU(3)_R\times U(1)_A/SU(3)_V$ is given by the 
$U(3)$ matrix $U=V \Sigma$. The quark mass matrix $M=\diag
(m,m,m_s)$ explicitly breaks the chiral symmetries of QCD, and 
enters the effective theory as a spurion transforming as 
$(3, \bar 3)_{(0,-2)} \oplus (\bar 3, 3)_{(0,2)}$.

The chiral Lagrangian introduced in \cite{Casalbuoni:1999wu} and elaborated 
 in \cite{Son:1999cm,Son:2000tu,Bedaque:2001je} is an expansion in quark mass, 
derivatives, and strong coupling $\alpha_s$. Generically it takes the 
form 
\be
{\cal L}\sim \mu^2 \Delta^2\; 
  \hat{\cal L}\left(
      \frac{\partial^2}{\Delta^2}, 
      \frac{M^2}{\mu\Delta}, 
      \frac{M^2}{\mu^2}, 
      \alpha_s, 
      \frac{\Delta^2}{\mu^2}\right)\ ,
\eqn{xpt}
\ee
where $\Delta=\Delta_3$. To date the theory has been discussed expanding 
the dimensionless function $\hat {\cal L}$ to include all terms of order 
$(\partial^2/\Delta^2)$, $(m_s^2/\mu\Delta)^2$, and $ (M^2/\mu^2)$; the  
Lagrangian to this order is
\ber
{\cal L} &=& {f^2_{\pi}\over{4}} \Tr 
\left( D_0 \Sigma D_0 \Sigma^{\dag} - v^2 \Tr \vec{\nabla} \Sigma \cdot 
  \vec{\nabla}\Sigma \right)\cr
 & & \mbox{}+ 
   \frac { f^2_{\eta^{'}}}{32}
   \left( \partial_0 V \partial_0 V^\dagger - v^2
   \vec{\nabla}V\cdot \vec{\nabla} V^\dagger  \right) \nonumber \\[0.1cm]
 & & \mbox{}+  
  a_3 V^\dagger\left[ (\Tr M \Sigma)^2 -
  \Tr M \Sigma M \Sigma  \right]+\text{ h.c.}
\eqn{lagrangian}
\eer 
The covariant derivatives in \eq{lagrangian} contain the Bedaque-Sch\"afer 
term \cite{Bedaque:2001je} that acts as a dynamically generated chemical 
potential for flavor symmetries,
\be
D_0 \Sigma = \partial_0 \Sigma 
    + i \left[M^2/2 \mu,\Sigma\right] .
\eqn{mueff}
\ee
By making use of the effective theory introduced by Hong
\cite{Hong:1998tn}, all of the parameters in \eq{lagrangian} have 
been computed to zeroth order in $\alpha_s$  
\cite{Son:1999cm,Son:2000tu,Schafer:2001za}.  The results
we will use in this paper are
\begin{eqnarray}
f^2_{\pi} = \frac{21-8 \ln 2}{18} \frac{\mu^2}{2 \pi^2}\ ,\qquad
a_3 = \frac{3}{4\pi^2}\left(\Delta_3\right)^2
\ .
\end{eqnarray}

The $a_3$ term in \eq{lagrangian}, which gives rise to the meson masses, 
can be rewritten as $2 a_3 V\,\Tr \tilde M \Sigma +\text{h.c.}$ 
where $\tilde M = \diag(m m_s, m m_s,m^2)$. It is therefore apparent 
that there exists a meson with square mass proportional to 
$ m^2 \Delta^2/\mu^2$, corresponding to fluctuation in the $\{3,3\}$ 
component of $V \Sigma$, which is exceedingly light.  A major point 
of this paper is that it does not make sense to keep a term proportional 
to $m^2$ while dropping ones proportional to $\alpha_s m_s^2$, given 
that $(m/m_s)^2 \simeq 1 \times 10^{-3}$. 
There are various small parameters that control the chiral expansion
in the CFL phase. Defining
 $\left(m_s^2/\mu\Delta_3\right)^2\equiv \delta$ and
 $m^2/\mu^2\equiv\epsilon$, we take $m_s m/\mu^2\sim \delta$ and
 $\alpha_s m_s^2/\mu^2 \sim \epsilon$ with
 $\delta^2\ll\epsilon\ll\delta$, expanding the Lagrangian to
 $O(\delta)$  and $O(\epsilon)$.

  
 Fig.~1 shows the leading order graphs that contribute to the $O(M^2)$ 
terms in the chiral effective lagrangian. Fermion lines in the graph 
on the left correspond to the anomalous quasi-particle propagators in 
the CFL phase and depend on the gaps $\Delta_{3,6}$. The four-fermion
vertex corresponds to a chirality changing four-fermion interaction
of order $O(g^2M^2)$ \cite{Schafer:2001za}. The Feynman diagram on 
the left in Fig.~1 gives rise to the $a_3$ term in in the low energy 
Lagrangian \eq{lagrangian}, as well as the new term
\ber
\delta{\cal L}_6 &=&- a_6\, V^\dagger
  \left[(\Tr M \Sigma)^2 + \Tr M\Sigma M \Sigma \right] + {\rm h.c.}\ , \\
  a_6 &=& \frac{3}{8\pi^2} \left(\Delta_6\right)^2 ,
\eqn{adef}
\eer
which is proportional to $\alpha_s$, given the relation \eq{dsix}
\footnote{The power of $g^2$ in the vertex of Fig.~1 is canceled by
the infrared logarithm $(\ln\Delta/\mu)^2\propto 1/g^2$.}. 
Perturbative corrections to the Feynman diagram in Fig.~1 that are not 
summed by the gap equation contribute at $O(\alpha_sm m_s\Delta^2)$. We 
neglect these terms in this work. We also neglect the contribution from 
the graph on the right of Fig.~1 which  is of order $O(m_s^2\Delta^4/
\mu^2)$, which is exponentially small as compared to the terms we have 
kept.

The contributions to the vacuum energy proportional to  $a_3$ and $a_6$ 
are opposite in sign, reflecting the fact that one gluon exchange is 
attractive in the color triplet channel, and repulsive in the sextet 
channel. Because of the minus sign in $\delta\CL_6$ one finds that the 
$O(\alpha_s m_s^2)$ contributions to the square of the meson masses 
are negative, and can alter the vacuum alignment.

\begin{figure}[t]
\centering{
\psfrag{x}{${\mathrm R}$}
\psfrag{y}{${\mathrm L}$}
\psfrag{d}{$\Delta$}
\psfrag{w1}{${\mathrm g^2 M^2}$}
\psfrag{w2}{${\mathrm g^2 M^2}$}
\psfrag{p}{$+$}
\epsfig{figure=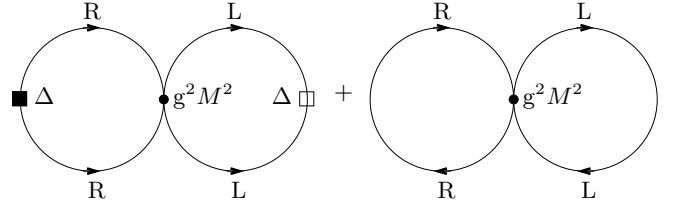, width=.48\textwidth}
 }
 \caption{Feynman diagrams corresponding to the $g^2 M^2$ 
 terms in the effective potential. The 
open and solid squares denote insertions of the gap, and 
the dot is an effective vertex which describes chirality 
changing quark-quark scattering amplitudes. }
 \label{fig1}
 \end{figure}

To analyze the vacuum alignment, we consider free energy density of
the homogeneous medium arising from $\CL + \delta\CL$:
\ber
\Omega &=& -\frac{f^2_{\pi}}{4} 
  \Tr \Bigl\vert \left[M^2/2 \mu,\Sigma\right]\Bigr\vert^2 
  \nonumber\\[0.1cm]
 && \mbox{}-
  {2 (a_3 + a_6)} V\,
  \Tr \tilde{M}\Sigma + \text{h.c.} \nonumber \\[0.1cm]
 && + 
  2 a_6 V^\dag\left( \Tr M  \Sigma\right)^2  + \text{h.c.}
\eqn{freeenergy}
\eer

Neglecting the new $a_6$ contributions, one finds that $\Omega$ is
minimized by the vacuum configuration 
\be
V=1\ ,\quad \Sigma=
\begin{pmatrix}
1 & 0 & 0 \\ 
0 & \ \cos \theta & i \sin \theta \\
0 & i \sin \theta & \ \cos \theta
\end{pmatrix}
\equiv \Sigma_K\ ,
\ee
where the angle $\theta$ is satisfies
\be
\cos \theta =\text{min}\left[1,c_0\right]\ ,\quad
c_0\equiv \frac{16 m a_3}{m_s^3}\left(\frac{\mu}{f_\pi}\right)^2\ ,
\eqn{c0def}
\ee
where we have dropped $O(m/m_s)$ corrections to $c_0$. This is 
the kaon condensed phase discussed in 
\cite{Schafer:2000ew,Bedaque:2001je,Kaplan:2001qk}.
Note that if we neglect the $a_6$ contribution, then  since 
$\tilde M_{33} = m^2$ there exists a very light excitation 
$\varphi$ about $\Sigma_K$, corresponding to 
\be
 V\Sigma = 
  \begin{pmatrix}
   1 &   & \cr
     & 1 & \cr
     &   & e^{i\varphi}
\end{pmatrix}\, \Sigma_K\ .
\ee
or
\be
V = e^{i\varphi/3}\ ,\quad \Sigma = 
\begin{pmatrix}
 e^{-i\varphi/3} & & \cr
      &  e^{-i\varphi/3} & \cr
      &      & e^{2i\varphi/3}
\end{pmatrix}\,\Sigma_K\ .
\eqn{phivac}
\ee
The $\varphi$ excitation has a mass $M_\varphi^2\propto (a_3
m^2/\mu^2)$ in the absence of kaon
condensation, which is very small.    Therefore it is in this direction that we expect an 
instability to arise when the $a_6$ operator is included, which
contributes $\delta M_\varphi^2\sim -(a_6 m_s^2/\mu^2)$. With the 
ansatz \eq{phivac}, we can expand  the free energy to first order 
in $\alpha_s $ and $m/m_s$, with $\Omega=\Omega_0+\Omega_1+\ldots$, 
where
\ber
\Omega_0 &=& -\frac{f_{\pi}^2   m_s^4  
  \sin^2 \theta}{8 \mu^2}
-  4 m m_s a_3 \cos\theta \cr 
\Omega_1&=& -4\left( a_3 m^2 \cos\theta -a_6 m_s^2
  \cos^2\theta\right)\cos\varphi\ .
\eer
We are neglecting contributions to the free energy  of order $m m_s 
a_6$, $m^2 m_s^2$ and smaller. 

\begin{figure}[t]
\centerline{\includegraphics[height=2.0in]{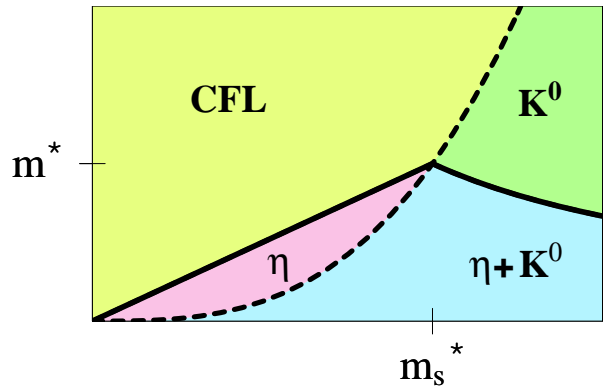} }
\caption{The phase diagram as a function of light quark mass $m$ and
  strange quark mass $m_s$. The phases marked CFL, $K^0$, $\eta$,
  and $\eta+K^0$ are respectively the CFL phase without meson
  condensation, with kaon condensation, with $\eta$ condensation, 
  and with both $\eta$ and $K^0$ condensation.  Phase transitions 
  are represented by a solid line if first order, a dashed line if second 
  order. The location of the tetracritical point is $\{m_s^*,m^*\}$, given 
  in \eq{mmsstar}; including subleading effects causes it to split into 
  two tricritical points.}
\label{pdiagram}\end{figure}
\hspace{1.25in}

Minimizing $\Omega_0$ with respect to $\theta$ determines the $K^0$ 
condensate. The subleading contribution to the free energy, $\Omega_1$, 
has a negligible effect on the kaon condensate, which is still given by
equ.~(\ref{eq:c0def}), but is the leading contribution to the potential 
for the angle $\varphi$. Minimizing with respect to $\varphi$ yields
\be
\cos\varphi = \text{Sign}\left[a_3 m^2-a_6 m_s^2\cos\theta\right]\ .
\eqn{phicon}
\ee
For brevity we will refer to the phase with $\cos\varphi=-1$ as an 
``$\eta$'' condensate, although the mode that condenses is actually 
a linear combination of $\eta$ and $\eta'$.  (Note that in this phase
$V = e^{i\pi/3}$; since $V$ has $U(1)_A$ charge equal to four, this
phase is equivalent to a strong $CP$ angle $\bar\theta_{QCD}=\pi/2$.)
Thus we see that we 
have four different phases, depending on the quark masses, corresponding 
to whether or not the kaon and/or the $\eta$ condense. 

The symmetry structure of the four phases is as follows.  The ordinary
CFL phase (neglecting isospin breaking) possesses an $H=SU(2)\times
U(1)_Y\times P$ symmetry, where $P$ is parity, under which
$\Sigma\to\Sigma^\dagger$ and  $V\to V^\dagger$. Neutral kaon
condensation by itself breaks $H$ down to $U(1)_{\text{em}}\times P'$, 
where $P'$ is a combination of ordinary parity and a discrete hypercharge 
rotation $\text{exp}(i\pi \hat Y)$, and $U(1)_{\text{em}}$ is the
electromagnetic group.  The $\eta$ condensate by itself breaks the 
discrete parity symmetry $P$, but no continuous symmetries \footnote{If 
$\langle \bar q_L \bar q_L q_R q_R\rangle \sim V\Sigma$ were the only 
order parameter, then the $\eta$ condensed phase with $\phi=\pi$ and 
$\theta=0$ would not break parity, since $V\Sigma$ would equal its 
hermitean conjugate (see \eq{phivac}). However, it is possible to 
construct a twelve fermion order parameter $\langle \left(q_L^4
q_R^2\right)\left(\bar q_L^2 \bar q_R^4\right)\rangle$ which
transforms as $(1,1)_{(0,4)}\sim V$.  Therefore $V=e^{i\pi/3}$
serves as an order parameter for parity violation in the $\eta$
condensed phase.}.   The phase with both $\eta$ and $K^0$
condensation breaks $H$ to $U(1)_{em}$.   

 The phase diagram is easily constructed by means of
Eqs.~(\ref{eq:c0def},\ref{eq:phicon}).  The phase boundary for kaon
condensation is second order and is described by an equation of the
form $m\propto m_s^3$;  it is unaffected by the existence or absence 
of $\eta$ condensation at the order in $\alpha_s$ and $m/m_s$ that 
we are working. The phase boundary for $\eta$ condensation is first 
order, as one might expect for the breaking of a discrete
symmetry. However, to the order we work the $\varphi$ angle does not
encounter a free energy barrier at the phase transition, and so it is
possible that higher order corrections could render the phase transition
second order.
It is described by a line $m\propto m_s$ in the region without kaon
condensation, and the curve $m\propto 1/m_s$ in the kaon condensed
region. The phase diagram  shown in Fig.~\ref{pdiagram}. At this 
order there is a tetra-critical point, which is separated into two 
tricritical points when higher order corrections are included. 
The coordinates of this critical point are $(m_s^*,m^*)$
given by
\ber
m^{*} &=&
 \frac{2 (\ln 2)^{3/2}}{9 \sqrt{3}\pi}\frac{\mu}{f_{\pi}} \Delta_3
 \left(\frac{\alpha_s}{4 \pi}\right)^{3/4} \cr
m_s^*&=&
 \frac{2 (\ln 2)^{1/2}}{\sqrt{3} \pi}\frac{\mu}{f_{\pi}} \Delta_3
 \left(\frac{\alpha_s}{4 \pi}\right)^{1/4} 
 \eqn{mmsstar}
\eer
If the baryon chemical potential is very large we can use 
perturbation theory in order to estimate the location of the 
tetracritical point. If we assume that there are only three flavors of
quark, and that $\alpha_s(m_\tau)=0.353$ as in the real world, then at
 $\mu=10^{10}$ GeV we have $\alpha_s/\pi=0.009$,
and the critical values
\be
\mu= 10^{10}\,\text{GeV}:\quad 
m^*=320\, \text{keV}\, \quad m_s^*=89 \,\text{MeV}\ .
\ee
In the regime $\mu\simeq 0.5$ GeV, relevant to the physics
of neutron stars, the coupling is large and perturbation
theory is not reliable. If we assume that $\alpha_s/\pi
\simeq 1$, $\Delta\simeq 100$ MeV, and $f_\pi\simeq 90$ 
MeV we find 
\be
\mu= 500\,\text{MeV}:\quad 
m^*\simeq 4.6\, \text{MeV}\, \quad m_s^*\simeq 120 \,\text{MeV}\ .
\ee
 It 
is interesting to note that the values of the critical 
masses are remarkably close to
the physical values, particularly in the strong coupling example.
 As a consequence, we expect
that 
the region of the phase diagram explored in this work and displayed in
Fig.~2 
may well  be relevant to the structure of CFL matter in 
neutron stars. 

 We emphasize that if CFL quark matter is the true ground
state of quark matter at large baryon density then our 
study provides a rigorous analysis of the phase diagram 
for small quark masses in the regime $m<m_s$. As such, 
it provides the correct starting point for an exploration of 
the phase structure for larger values of the quark masses,
and an important guide for any attempt to numerically
simulate QCD in the regime of very large baryon density 
and low temperature.

\begin{center}
\large{ \textbf{Acknowledgments}}
\end{center}
We wish to thank Sanjay Reddy and Dam T. Son for helpful discussions along the way.  
The work is supported in part by the US Department of Energy grant
DE-FG02-00ER41132 (A.K. and D.B.K) and DE-FG-88ER40388 (T.S.).

\bibliography{cfl}
%

\end{document}